# Detection of extensive air showers with the NEVOD-EAS cluster type detector

I. A. Shulzhenko, M. B. Amelchakov, N. S. Barbashina, A. G. Bogdanov, N. E. Fomin,
S. S. Khokhlov, N. N. Kamlev, R. P. Kokoulin, K. G. Kompaniets, O. I. Likiy, V. V. Ovchinnikov,
A. A. Petrukhin, V. V. Shutenko, I. I. Yashin
*National Research Nuclear University MEPhI (Moscow Engineering Physics Institute), 115409 Moscow, Russia*

A. Chiavassa
*INFN, 10125 Torino, Italy and National Research Nuclear University MEPhI (Moscow Engineering Physics Institute), 115409 Moscow, Russia*

O. Saavedra
*Dipartimento di Fisica dell' Università degli Studi di Torino, 10125 Torino, Italy and National Research Nuclear University MEPhI (Moscow Engineering Physics Institute), 115409 Moscow, Russia*

G. Mannocchi, G. Trinchero
*INAF, 10133 Torino, Italy*

A new cluster type shower array NEVOD-EAS is designed for estimating the size, axis position and arrival direction of extensive air showers registered by the Cherenkov water detector NEVOD and coordinate detector DECOR (Moscow, Russia). In 2015-2016, the central part of the array was deployed and started its operation. It includes 4 independent clusters of scintillation detector stations located around the NEVOD-DECOR experimental complex on the area of about $10^4$ m$^2$. This article presents the results of studying amplitude and timing characteristics of the array clusters which are critical for EAS parameters reconstruction, as well as the examples of registered events.

## I. INTRODUCTION

At the Experimental Complex NEVOD (Moscow, Russia) which includes Cherenkov water detector NEVOD [1, 2] and coordinate detector DECOR [3], the muon bundles are studied in a wide range of zenith angles and energies using the method of local muon density spectra (LMDS) [4]. The LMDS technique has low accuracy of primary particle energy estimations ($\sigma(\lg E_0) \sim 0.4$) due to the contribution of extensive air showers registered at different distances from the setup to the events with a fixed local muon density.

A new shower array of the Experimental Complex NEVOD (NEVOD-EAS) which is now being constructed will allow independent estimation of EAS arrival direction, size, axis position in the energy range from $10^{15}$ to $10^{17}$ eV using classical technique and, therefore, will enable to calibrate the LMDS method and to increase its accuracy.

## II. NEVOD-EAS SHOWER ARRAY

NEVOD-EAS shower array is organized by the cluster principle and its first stage will include 12 independent clusters located around the experimental complex NEVOD-DECOR on the roofs of MEPhI laboratory buildings and on the ground. The array clusters consist of 16 scintillation counters of EAS electron-photon component particles. Counters of the cluster are combined in 4 detector stations (DS) placed in the vertices of a rectangle with typical dimensions of about $15\times15$ m$^2$. The distance between clusters is $\sim$ 50 m. The array total area will reach $2\times10^4$ m$^2$.

Scintillation counter of the array consists of plastic scintillator NE102A ($800\times800\times40$ mm$^3$) and photomultiplier Philips XP3462 (3″ hemi-spherical photocathode, 8 dynodes) for arrival time and particle density measurements installed inside the pyramidal housing. One additional PMT can be installed inside the housing. Counters were previously used in the KASCADE-Grande [5] (Germany) and EAS-Top [6] (Italy) experiments. The upper limit of the counter dynamic range is adjusted to 100 particles (vertical equivalent muons, VEM). The light collection nonuniformity of counters does not exceed 25%. The results of detailed study of characteristics of PMTs, scintillators and counters are presented in [7-9].

Detector station includes 4 counters. Its total area is 2.56 m$^2$. One of 4 counters is equipped with an additional PMT with a gain of 90 times lower than the other ("standard") PMTs and provides wide linearity range at high particle densities ($\sim$10000 particles/m$^2$). The description of the DS is presented in [10].

Analog signals from the cluster detector stations are fed to the Local Post (LP) of primary data processing. It performs summation of standard PMT signals, signal digitizing, event selection according to intra-cluster triggering conditions (registration thresholds, coincidence multiplicity, time gate), event time stamping and data transfer to the array Central DAQ Post (CP).

Central Post synchronizes (with accuracy of 10 ns) and controls LP of the array, as well as stores experimental and monitoring data. The characteristics and operating principles of the LP and CP are described in [11].

In 2015-2016, the array central part including 4 clusters was created and started. Three clusters are located on the roofs of MEPhI buildings, the fourth one is on the ground. The central part area is about $10^4$ m$^2$. Photographs of the central part clusters are shown in Figure 1.

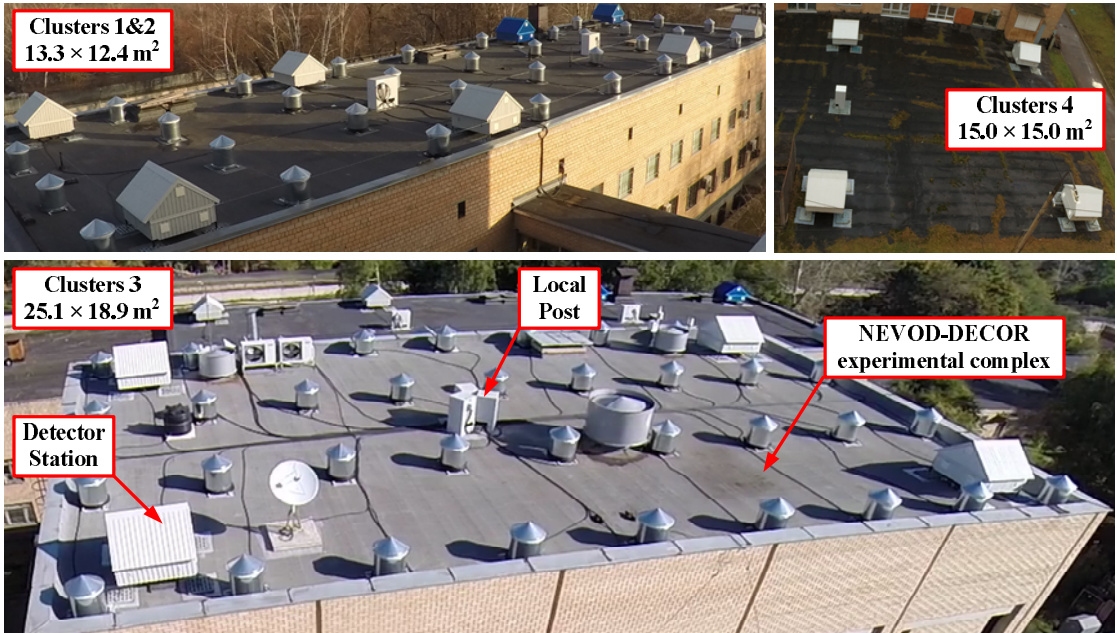

Figure 1: Clusters of the NEVOD-EAS array central part.

### III. CLUSTER CHARACTERISTICS

During the first test runs the characteristics (the responses of detector stations and their counters to 1 VEM, the responses of DS additional PMTs to 1 VEM and the accuracy of the DS hit time determination) critical for the correct reconstruction of registered EAS parameters were measured for all 4 clusters of the NEVOD-EAS array central part.

The typical response charge spectra of the DS (No.1, 1$^{st}$ cluster) and its counters measured in self-triggering mode are shown in Figure 2. Measurement time for the DS spectrum is 4 times longer than for the spectrum of each counter. The orange line shows the sum of the counters spectra. The most probable values (MPV) of the obtained "muon" peaks are indicated in the legend. A good accordance between the MPVs of the DS and summed spectra is observed. The MPV of the DS spectrum characterizes its response to 1 VEM and is used for calculating the number of registered particles.

In Figure 3, the ratio between the response charges of DS No.1 of the 1$^{st}$ cluster ($Q_{St}$) and its additional PMT ($Q_{Add}$) is plotted as a function of the DS response charge.

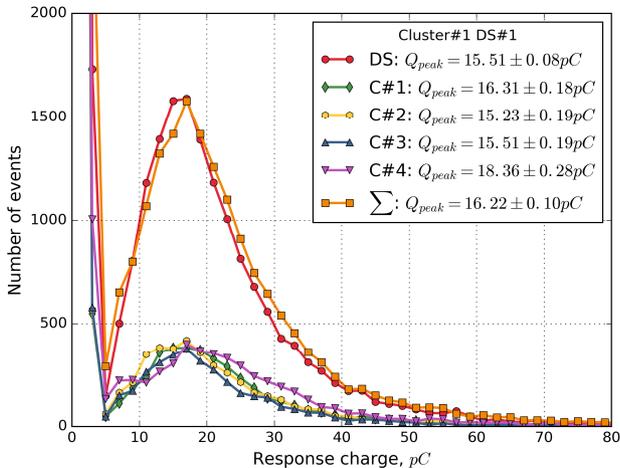

Figure 2: Response charge spectra for the DS No.1 of the 1$^{st}$ cluster and its counters.

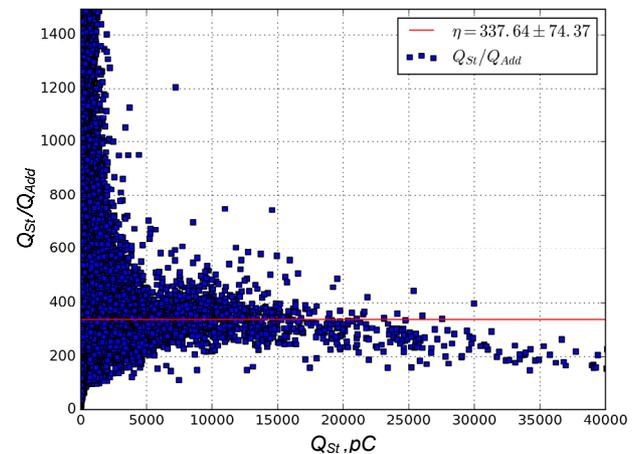

Figure 3: Ratio between the response charges of DS No.1 of the 1$^{st}$ cluster and its additional PMT as a function of the DS response charge.

It is seen that for $Q_{St}$ from 5000 to 20000 pC the ratio is constant. The mean ratio $\eta$ (red horizontal line) in this range is ~ 338. Using the MPV of the DS response charge spectrum (see Figure 2), the response of additional PMT to 1 particle VEM can be calculated. For the considered DS it is ~ 0.18 pC.

Estimation of the DS hit time determination accuracy was carried out at the 1$^{st}$ cluster. The DS No.2 was disconnected from the LP. Two counters of the DS No.1 were connected instead of the DS No.2. Thus, the DS No.1 was divided into 2 stations (1* and 2*) of 2 counters. In such cluster configuration, at the moment of EAS registration, both stations must be triggered almost simultaneously for any EAS arrival directions.

Measurement lasted 12 hours. The 4-fold coincidence of DS within 120 ns time gate was used as intra-cluster triggering condition. According to the obtained data, the hit time difference between DS 1* and 2* was calculated for all registered events.

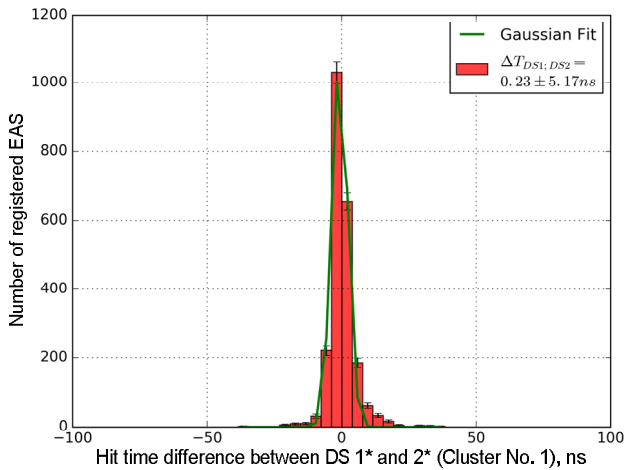

Figure 4: Distribution of registered EAS in the hit time difference between DS 1* and 2* of the 1$^{st}$ cluster.

Figure 4 shows the distribution of registered EAS in the hit time difference between detector stations 1* and 2*. The mean hit time difference is about 0.23 ns. That means that DS 1* and 2* were triggered almost simultaneously in all registered events. The rms value for this distribution is ~ 5.17 ns. So, the estimated hit time determination accuracy is about 3.7 ns.

## IV.    EAS DETECTION

As the NEVOD-EAS array clusters are deployed not in the same plane, but at different altitudes, a cluster approach to the registered EAS reconstruction has been developed. The arrival direction of extensive air showers is determined in the following way. The events in which not less than 3 clusters were triggered during the time interval of about 0.5 μs are selected. Such duration of the time interval is determined by the characteristic size of the array (~ 100 m). At the present stage, during the experimental runs a 4-fold coincidence of detector stations within a 120 ns time gate is used as an intra-cluster triggering condition. Then the arrival direction for each cluster is estimated in the assumption of flat EAS front.

Figures 5 and 6 show the distributions of EAS registered by the 1$^{st}$ cluster in the cosine of the arrival direction zenith angle and in the arrival direction azimuthal angle, correspondingly.

It is seen that the distribution in the cosine of EAS arrival direction zenith angle is in a good accordance with the expected power-law dependence of EAS flux on the zenith angle cosine $I(\theta) = I_0 \cdot \cos^n \theta$ with $n$ of about 8.5.

The distribution in the EAS arrival direction azimuthal angle is close to uniform.

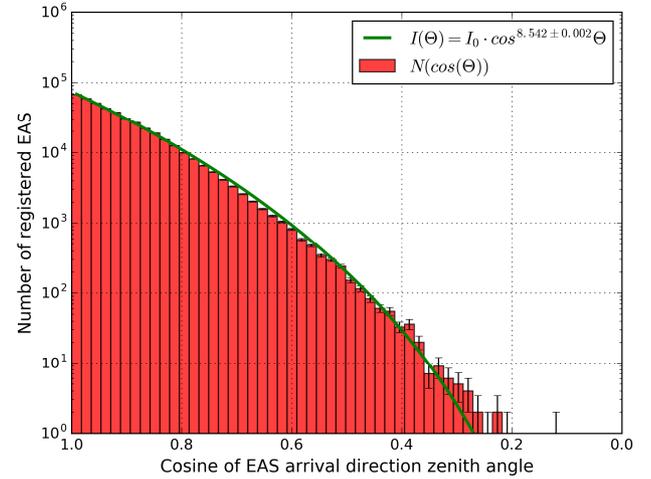

Figure 5: Distribution of EAS registered by the 1$^{st}$ cluster in the cosine of the arrival direction zenith angle.

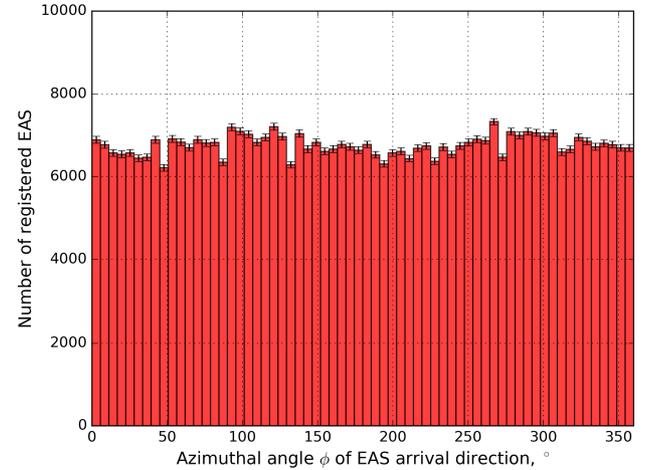

Figure 5: Distribution of EAS registered by the 1$^{st}$ cluster in the arrival direction azimuthal angle.

The resulting direction in the event is calculated by averaging of directions reconstructed according to the data of triggered clusters. The number of particles registered by the detector station is determined as the ratio between the DS response charge in the event and its response charge to 1 VEM measured during the test runs. The responses of the DS additional PMTs are accounted at high particle densities (> 100 particles/m$^2$).

Figure 7 shows the EAS reconstructed according to the data of NEVOD-EAS standard and additional PMTs.

Rectangles represent the locations of clusters. Squares are for the stations. Color corresponds to the number of particles $N_{St}$ (VEM) registered by standard PMTs of each DS. The numbers of particles registered by the additional PMTs $N_{Add}$ of detector stations are also noted. A good accordance between $N_{St}$ and $N_{Add}$ is observed at numbers of particles higher than ~ 130 particles/DS.

The angles of EAS arrival direction reconstructed according to the data of separate clusters (the hit times of their standard PMTs are accounted) are close to each other: zenith angles are about 20°±2° and azimuthal angles are in range from 266.5° to 292.2°.

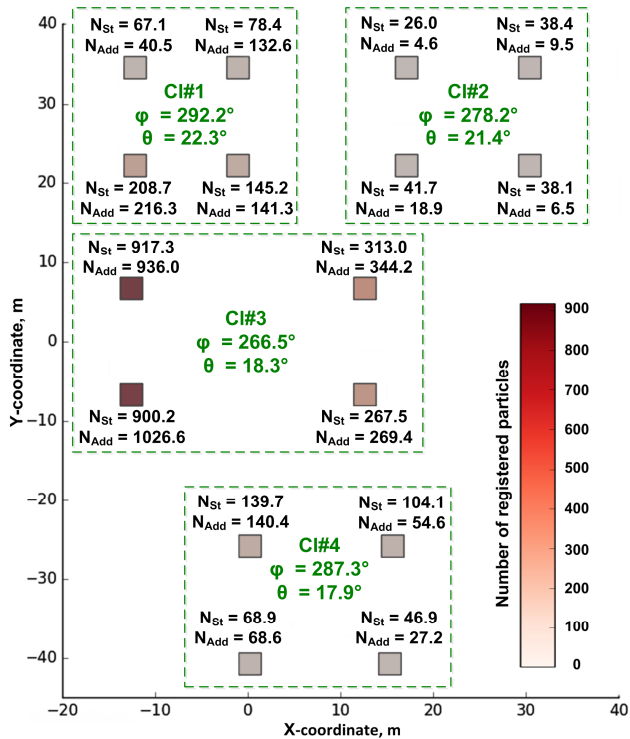

Figure 7: Visualization of the EAS registered by the central part of the NEVOD-EAS array.

## V. CONCLUSION

In 2015–2016, the central part of the NEVOD-EAS cluster type shower array was created and started. It includes 4 independent clusters deployed around the NEVOD-DECOR experimental complex at the area of about $10^4$ m$^2$.

The distributed structure of the NEVOD-EAS registering system allows deployment of the array detecting elements even at different altitudes and provides implementation of a newly developed cluster approach to the EAS parameters reconstruction. First tests and experimental runs have proved the possibility of using of such cluster approach.

### Acknowledgments

This work was performed at the Unique Scientific Facility "Experimental complex NEVOD" with the support of the Ministry of Education and Science of the Russian Federation (contract RFMEFI59114X0002, MEPhI Academic Excellence Project 02.a03.21.0005 of 27.08.2013) and the Russian Foundation for Basic Research (grant 16-29-13028-ofi-m).